\documentclass{article}

\def\xxinput#1{\input#1}

\xxinput{vsolj02.sty}

\usepackage[dvipdfmx]{graphicx}
\usepackage[comma,colon]{natbib}

\usepackage[OT2,T1]{fontenc}

\def\cite{\citealt}
\setcitestyle{aysep={}}

\newcounter{author}
\def\altaffilmark#1{$^{#1}$}
\def\altaffiltext#1{$^{#1}$\,}

\def\authorcount#1#2{{\refstepcounter{author}\label{#1}
                     \altaffiltext{\ref{#1}}{#2}}}

\begin{document}

\begin{center}

\title{Gaia19bxc: possible polar below the period minimum}

\author{
        Taichi~Kato\altaffilmark{\ref{affil:Kyoto}}
}
\email{tkato@kusastro.kyoto-u.ac.jp}

\authorcount{affil:Kyoto}{
     Department of Astronomy, Kyoto University, Sakyo-ku,
     Kyoto 606-8502, Japan}

\end{center}

\begin{abstract}
\xxinput{abst.inc}
\end{abstract}

   Gaia19bxc is a transient detected by the Gaia Photometric
Science Alerts Team\footnote{
  $<$http://gsaweb.ast.cam.ac.uk/alerts/alert/Gaia19bxc/$>$.
}.  The object was detected at 19.28 mag on 2019 May 9
and was reported as a ``variable Gaia source brightens by 2 mag''.
The object stayed bright for months after this
according to the Gaia light curve.
The object is located at
17$^{\rm h}$ 31$^{\rm m}$ 58\hbox{$.\mkern-4mu^{\rm s}$}468,
$+$27$^{\circ}$ 09$^\prime$ 36\hbox{$.\mkern-4mu^{\prime\prime}$}12
(J2000.0) has $BP$=20.570(188), $RP$=19.726(142)
and a parallax $\varpi$=0.565(772) mas \citep{GaiaEDR3}.
The American Association of Variable Stars
(AAVSO) Variable Star Index (VSX: \cite{wat06VSX})
classified it as a possible dwarf nova (UG:).
The Zwicky Transient Facility (ZTF: \cite{ZTF})
listed it as a suspected variable star \citep{ofe20ZTFvar}.
I analyzed the light variation using the ZTF public data\footnote{
  The ZTF data can be obtained from IRSA
$<$https://irsa.ipac.caltech.edu/Missions/ztf.html$>$
using the interface
$<$https://irsa.ipac.caltech.edu/docs/program\_interface/ztf\_api.html$>$
or using a wrapper of the above IRSA API
$<$https://github.com/MickaelRigault/ztfquery$>$.
}.

\begin{figure*}
\begin{center}
\includegraphics[width=16cm]{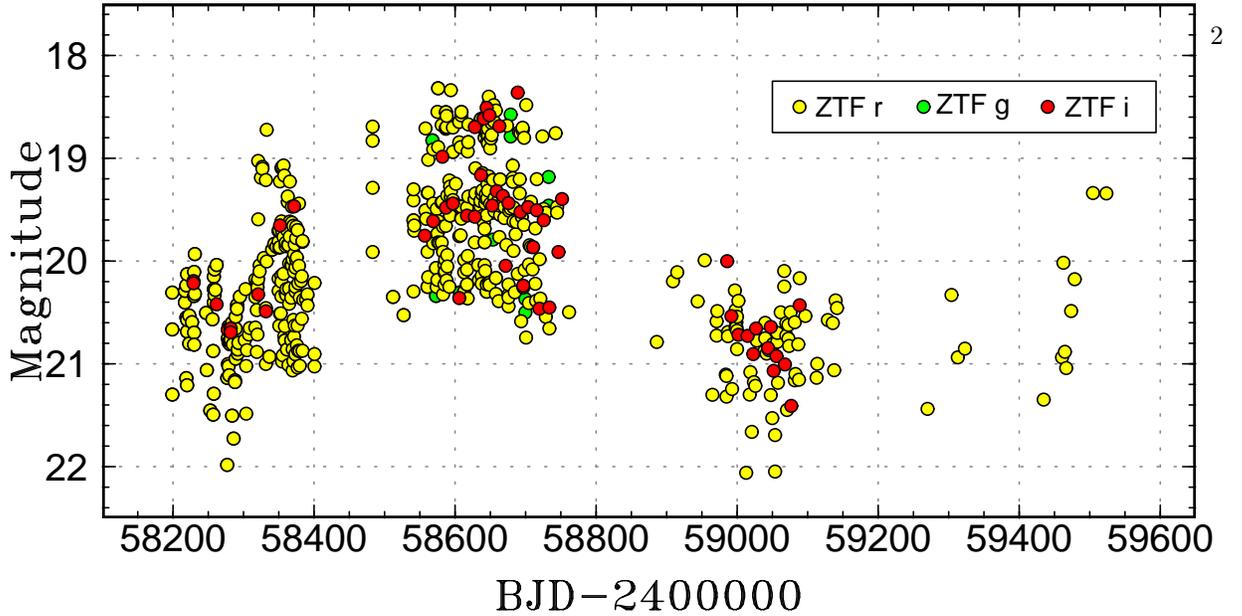}
\caption{
  Long-term ZTF light curve of Gaia19bxc.
}
\label{fig:long}
\end{center}
\end{figure*}

\begin{figure*}
\begin{center}
\includegraphics[width=16cm]{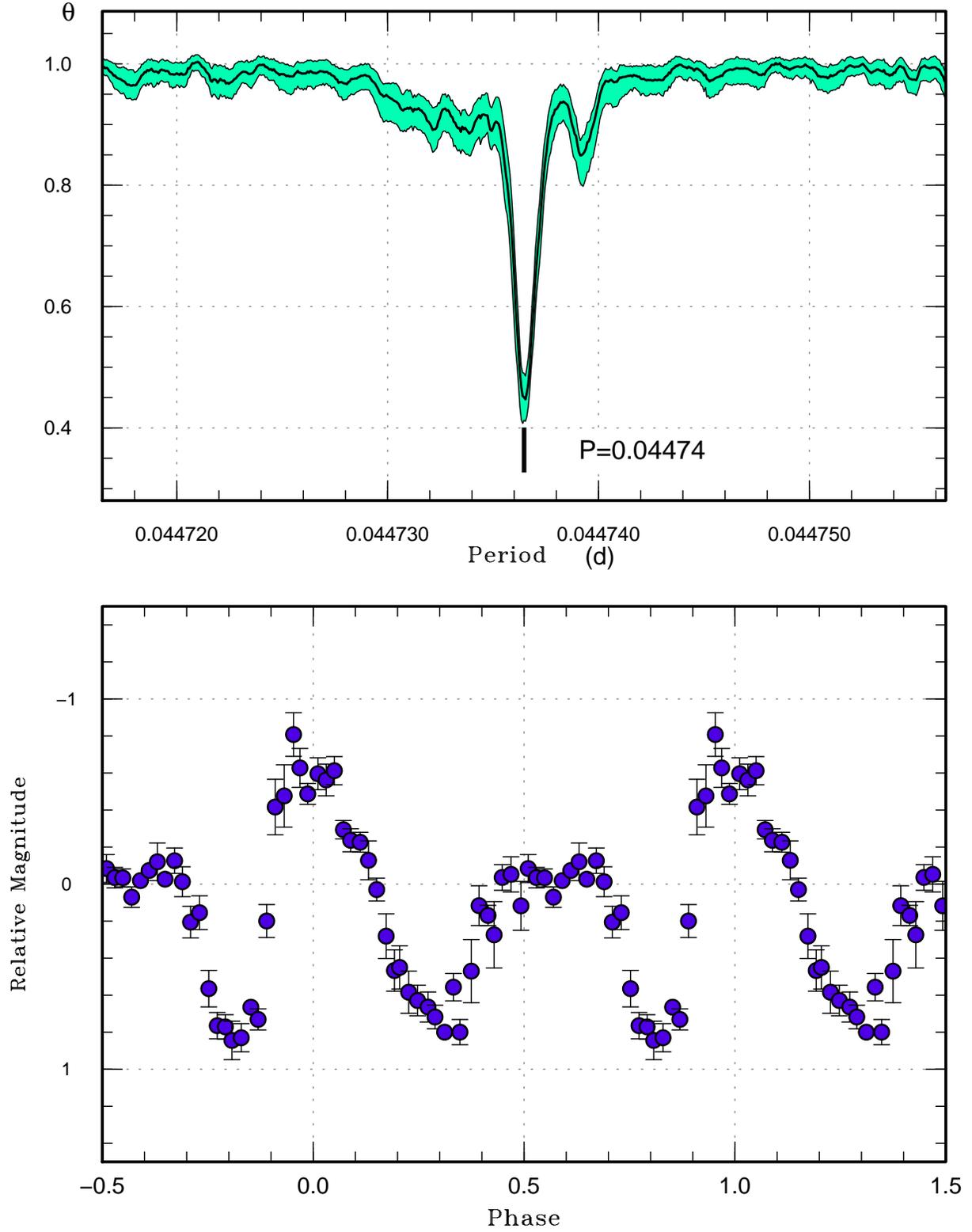}
\caption{PDM analysis of Gaia19bxc using the ZTF data.
  (Upper): PDM analysis.  The bootstrap result using
  randomly contain 50\% of observations is shown as
  a form of 90\% confidence intervals in the resultant 
  $\theta$ statistics.
  (Lower): mean profile.  1$\sigma$ error bars are shown.
}
\label{fig:pdm}
\end{center}
\end{figure*}

\begin{figure*}
\begin{center}
\includegraphics[width=16cm]{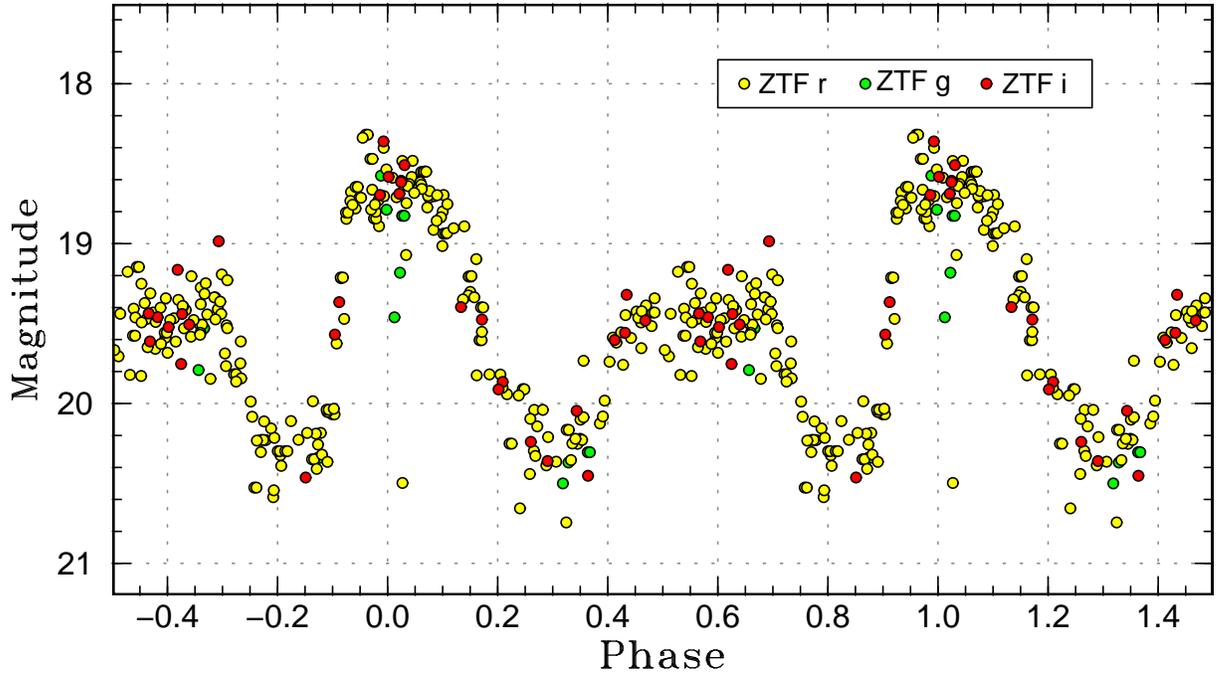}
\caption{The light curve of Gaia19bxc in high state
  (BJD 2458500--2458800) using
  the ZTF data folded by the 0.04473647(3)~d
  The zero epoch was chosen as BJD 2458563.005.
}
\label{fig:phase}
\end{center}
\end{figure*}

\begin{figure*}
\begin{center}
\includegraphics[width=16cm]{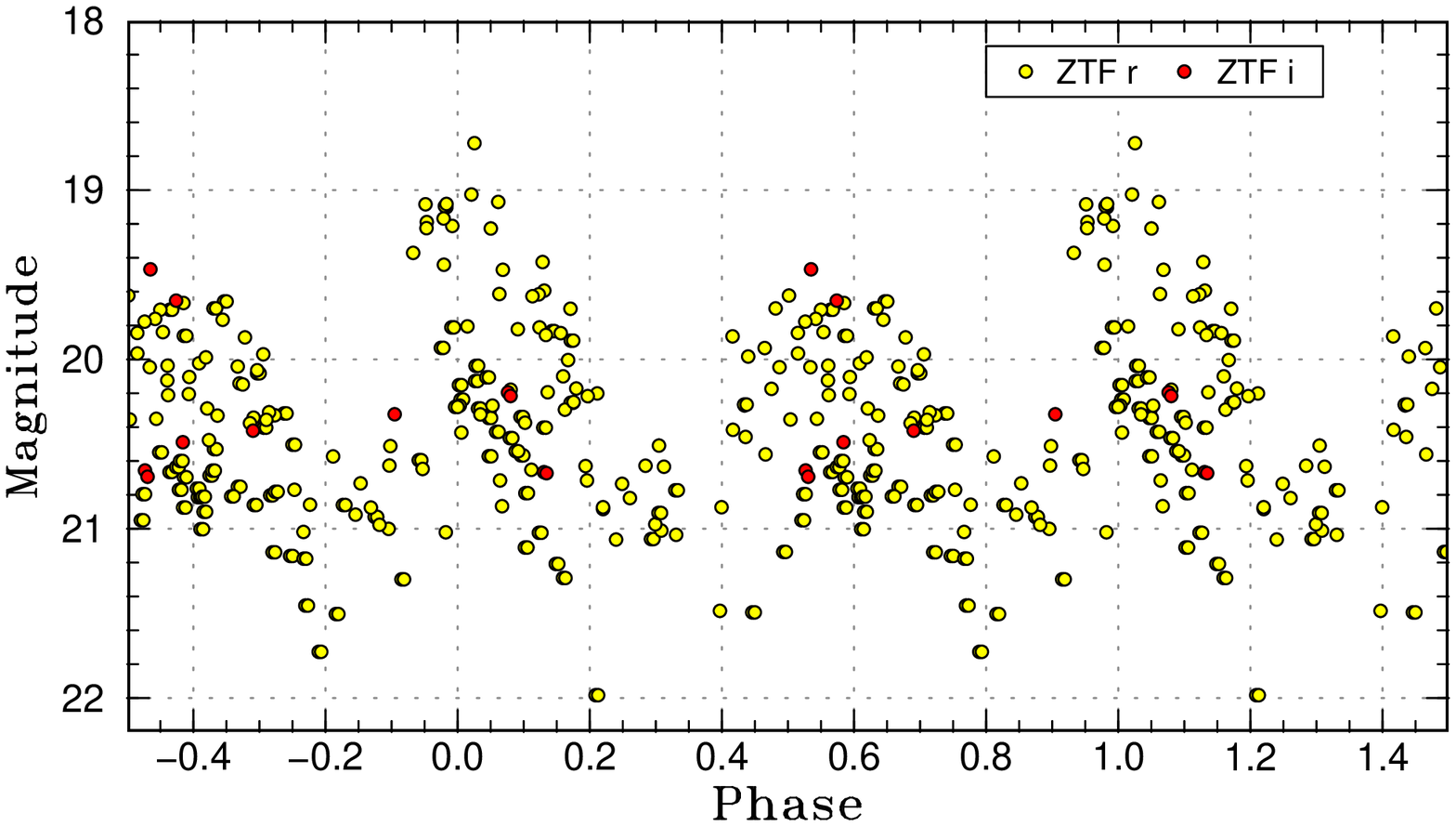}
\caption{The light curve of Gaia19bxc in intermediate
  (before BJD 2458400).  The period and epoch are
  the same as in figure \ref{fig:phase}.
}
\label{fig:phase2}
\end{center}
\end{figure*}

   The long-term variation is shown in figure \ref{fig:long}.
The object apparently shows high and low states and
the variation is not that of a dwarf nova.
A phase dispersion minimization (PDM: \cite{PDM}) analysis
of the ZTF data after removing the global trend
using locally-weighted polynomial regression (LOWESS: \cite{LOWESS})
yielded a period of 0.04473647(3)~d (figure \ref{fig:pdm}).
There were two maxima of different amplitudes in one
cycle.  The amplitude of the variation was very large,
reaching 2.0~mag.  The phased light curve of the bright
state is shown in figure \ref{fig:phase}.  The phased
light curve for the intermediate state is shown in
figure \ref{fig:phase2}.  The profile was almost the same
between different states and the colors between the different
ZTF bands were almost zero.  The phased light curve of
the low state could not be examined due to the small number
of observations.  The SDSS colors \citep{SDSS16} at four
different epochs had $u-g$ of $+$0.1 to $+$1.6 and $i-z$
of $+$0.0 to $+$0.9.  The $u-g$ values seem to be
redder than many polars, although individual measurements
were apparently affected by large-amplitude, short-term
variations.  The large variation was already
apparent in the SDSS data ($g$=21.56--23.04).

   No known type of pulsating variables has such
a large amplitude and a short period at the same time.
The flat maximum is not characteristic to a short-period
pulsating variable.
The ZTF colors (almost zero color indices) suggest
a blue object like a cataclysmic variable (CV).
The period is then either the spin period of an intermediate
polar (IP) with a very large amplitude as in Romanov V48
\citep{kat22romanov48}, the spin period of a white dwarf
pulsar like AR Sco \citep{mar16arsco,sti18arsco} and
candidates \citep{kat21j185139,kat21j205543rev},
or the orbital period ($P_{\rm orb}$) of a polar.
The presence of high and low states is also compatible
with a polar.
I could not find a longer period (corresponding to
$P_{\rm orb}$ for an IP) as in Romanov V48.
If the object is indeed a polar, this is very unusual.
The $P_{\rm orb}$ is far shorter than the ``period minimum''
of CVs \citep{kol93CVpopulation,gan09SDSSCVs,kni11CVdonor,
kat22stageA},

   There are several CVs below the period minimum.
One notable case is OV Boo with 
$P_{\rm orb}$ of 0.046258~d
\citep{lit07j1507,pat08j1507,uth11j1507,pat17ovboo}.
OV Boo has a large space velocity (against the Solar system)
and is considered to be a population II CV \citep{pat08j1507}.
It is possible that OV Boo is on the standard evolutionary
sequence of population II CVs.  There is no good distance
estimate for Gaia19bxc and a possibility for
a population II CV is not readily excluded.

   There are several hydrogen-rich CVs below the period
minimum.  The best known example are
V485 Cen with $P_{\rm orb}$=0.040995~d \citep{aug96v485cen},
EI Psc with $P_{\rm orb}$=0.044567~d \citep{tho02j2329}
CRTS J174033.4$+$414756 with $P_{\rm orb}$=0.045048~d
\citep{Pdot5} and
SBS 1108$+$574 with $P_{\rm orb}$=0.03845~d
\citep{lit13sbs1108}.  EI Psc is renowned to have
a hot, luminous secondary for this $P_{\rm orb}$
\citep{tho02j2329}.  These objects are considered to
have a secondary star with a stripped evolved core
and are also referred to as EI Psc stars.
It has been theoretically shown that CVs with $P_{\rm orb}$
shorter than the period minimum (including AM CVn stars)
can be formed if the core of the secondary star starts
evolving in the early phase of their evolution as CVs
\citep[see e.g.][]{nel10amcvnformation,pod03amcvn}.

   The lack of systems below the period minimum
among hitherto discovered polars is striking considering
the number of EI Psc-type objects  for non-magnetic CVs
[see figure 3 in \citet{bel20polarbraking}],
although there appears to be no specific reason why there
is no EI Psc-like object having a strongly magnetized
white dwarf.
1RXS J035410.4$-$165244 = RBS 490 was suggested to be a polar
with a possible $P_{\rm orb}$ of 46~min = 0.032~d by
a radial-velocity study \citep{tho06rbs0490,har15polarWISE}.
The most recent observation suggested a photometric
$P_{\rm orb}$ of 0.0703~d \citep{jos22rbs490j0759}.
I, however, could not find a significant period around
these periods in this system using the ZTF data.
The He II line is weak for a polar in the spectrum and
the field strength appears to be low even if it is indeed
a polar \citep{jos22rbs490j0759}.

   If Gaia19bxc is indeed confirmed to be a polar by
polarimetry, X-ray observations and spectroscopy,
it will be the first object below the period minimum.
Since the distance is poorly known, it is not easy
to tell whether this object has a secondary with
an evolved core from photometric data.  Such a secondary,
however, might explain the relatively red $u-g$ color
(the $u-g$ color of EI Psc is $+$0.8).
If this possibility is indeed the case, Gaia19bxc could become
a highly magnetized exotic ultracompact binary during its
secular evolution.

\section*{Acknowledgements}

This work was supported by JSPS KAKENHI Grant Number 21K03616.
The author is grateful to the ZTF team
for making their data available to the public.
We are grateful to Naoto Kojiguchi for
helping downloading the ZTF data.
This research has made use of the AAVSO Variable Star Index
and NASA's Astrophysics Data System.

Based on observations obtained with the Samuel Oschin 48-inch
Telescope at the Palomar Observatory as part of
the Zwicky Transient Facility project. ZTF is supported by
the National Science Foundation under Grant No. AST-1440341
and a collaboration including Caltech, IPAC, 
the Weizmann Institute for Science, the Oskar Klein Center
at Stockholm University, the University of Maryland,
the University of Washington, Deutsches Elektronen-Synchrotron
and Humboldt University, Los Alamos National Laboratories, 
the TANGO Consortium of Taiwan, the University of 
Wisconsin at Milwaukee, and Lawrence Berkeley National Laboratories.
Operations are conducted by COO, IPAC, and UW.

The ztfquery code was funded by the European Research Council
(ERC) under the European Union's Horizon 2020 research and 
innovation programme (grant agreement n$^{\circ}$759194
-- USNAC, PI: Rigault).

We acknowledge ESA Gaia, DPAC and the Photometric Science
Alerts Team (http://gsaweb.ast.cam.ac.uk/\hspace{0pt}alerts).
This research has made use of the VizieR catalogue access tool, CDS,
Strasbourg, France (DOI: 10.26093/cds/vizier). The original description 
of the VizieR service was published in 2000, A\&AS 143, 23.

\section*{List of objects in this paper}
\xxinput{objlist.inc}

\section*{References}

We provide two forms of the references section (for ADS
and as published) so that the references can be easily
incorporated into ADS.

\renewcommand\refname{\textbf{References (for ADS)}}

\newcommand{\noop}[1]{}\newcommand{\hyphalt}{-}

\xxinput{bxcaph.bbl}

\renewcommand\refname{\textbf{References (as published)}}

\xxinput{bxc.bbl.vsolj}

\end{document}